\documentstyle[epsf,twoside,fleqn,espcrc2]{article}

\newcommand{\AmS}{{\protect\the\textfont2
  A\kern-.1667em\lower.5ex\hbox{M}\kern-.125emS}}

\title{Gonihedric Ising Actions}

\author{D. A. Johnston\address{Mathematics Department, Heriot-Watt University, \\
        Edinburgh, EH14 4AS, United Kingdom}
        and
        Ranasinghe P. K. C. Malmini\address{{\it Permanent Address:} \\
	Department of Mathematics\\
	University of Sri Jayewardenepura\\
	Gangodawila, Sri Lanka.}
\thanks{Supported by Commmonwealth Fellowship SR0014}
}
       
\begin{document}

\begin{abstract}
We discuss a
generalized Ising action containing
nearest neighbour, next to nearest
neighbour and plaquette terms that has been suggested as a potential
string worldsheet discretization on cubic lattices
by Savvidy and Wegner. This displays both first and second order
transitions depending on the value of a ``self-intersection''
coupling as well as possessing a novel semi-global symmetry.
\end{abstract}

\maketitle

\section{INTRODUCTION}
Savvidy et al. \cite{1}
recently suggested a novel discretized
random surface theory, the so-called Gonihedric string,
\begin{equation}
S = {1 \over 2} \sum_{<ij>} | \vec X_i - \vec X_j | \theta (\alpha_{ij}),
\label{e4a}
\end{equation}
where the sum is over the edges of some triangulated surface,
$\theta(\alpha_{ij}) = | \pi - \alpha_{ij} |^{\zeta}$,
$\zeta$ is some exponent,
and $\alpha_{ij}$ is the dihedral angle between
neighbouring triangles with common link $<ij>$.
This definition of the action was
inspired by the geometrical notion
the linear size
of a surface, originally defined by Steiner.

In equ.(\ref{e4a}) it is the surface itself that is
discretized, rather than the space in which it is embedded.
An alternative approach
to discretizing the linear size is to
discretize the embedding space by
restricting the allowed surfaces to the plaquettes of a (hyper)cubic lattice.
This method
was applied by Savvidy and Wegner \cite{7,8,8a,8b},
who rewrote the resulting theory
as a generalized Ising model by using
the geometrical spin cluster boundaries
to define the surfaces.
The energy of a surface on a cubic
lattice is given explicitly
in the Savvidy-Wegner models by
$E=n_2 + 4 \kappa n_4$, where $n_2$ is the number
of links where two plaquettes meet at a right angle,
$n_4$ is the number of links where four plaquettes
meet at right angles, and $\kappa$ is a free
parameter which determines the relative
weight of a self-intersection of the surface.
In the limit $\kappa \rightarrow \infty$
the surfaces would be strongly self-avoiding,
whereas the opposite limit $\kappa \rightarrow 0$
would be that of phantom surfaces that could
pass through themselves without hindrance.
This should be contrasted with the standard 3D Ising
model with nearest neighbour
interactions where the surfaces are weighted by their
areas.
 
On a cubic lattice
the generalized ``Gonihedric'' Ising Hamiltonian which
reproduces the desired energy
$E=n_2 + 4 \kappa n_4$ contains nearest neighbour ($<i,j>$),
next to nearest neighbour ($<<i,j>>$) and round a plaquette ($[i,j,k,l]$)
terms
\begin{eqnarray}
H &=& 2 \kappa \sum_{<ij>}\sigma_{i} \sigma_{j} -
\frac{\kappa}{2}\sum_{<<i,j>>}\sigma_{i} \sigma_{j} \nonumber \\
&+& \frac{1-\kappa}{2}\sum_{[i,j,k,l]}\sigma_{i} \sigma_{j}\sigma_{k} \sigma_{l}.
\label{e1}
\end{eqnarray}
Such generalized Ising actions, or their
equivalent surface formulations, have quite complicated phase structures
for generic choices of the couplings \cite{9,9a,10}. The particular
ratio of couplings in equ.(\ref{e1}), however, is non-generic
and introduces a novel
symmetry into the model - it is possible to flip any plane
of spins at zero energy cost.

Although not a local gauge
symmetry, the flip symmetry of the model is intermediate between this
and a global symmetry.
This symmetry poses something of a problem when carrying out
simulations, as it means that a
simple ferromagnetic order parameter
\begin{equation}
M = \left< {1 \over L^3} \sum_i \sigma_i \right>.
\label{ord}
\end{equation}
will be zero in general, because of the layered
nature of the ground state. Even staggered
magnetizations would not do as the interlayer
spacing can be arbitrary.
It is possible, however, to
force the model into the ferromagnetic
ground state, which is equivalent
to any of the layered ground states, with a suitable choice of
boundary conditions on a finite lattice.
Fixed boundary conditions would do the job by penalizing
any flipped spin planes by a boundary term, but they carry
a higher price in finite size effects than the
customary periodic boundary conditions. A more elegant
solution is to fix any two internal planes of spins
in the lattice, whilst retaining the periodic boundary conditions.
This has the desired effect of picking out the ferromagnetic
ground state, whilst minimizing any finite size effects.
With fixed spin planes we can therefore still employ
the simple order parameter of equ.(\ref{ord}).

\section{ZERO-TEMPERATURE AND MEAN FIELD}

As the Gonihedric model is
a special case of the general action considered in \cite{9}
we can apply the methods used there
for both the zero temperature phase diagram and mean field
theory.
For the zero temperature
case this involves writing the full lattice
Hamiltonian as a sum over individual cube Hamiltonians
\begin{eqnarray}
h_c &=& \frac{\kappa}{2}\sum_{<i,j>} \sigma_{i} \sigma_{j} - \frac{\kappa}{4}  \sum_{<<i,j>> }\sigma_{i} \sigma_{j} \nonumber \\
&+&  \frac{1-\kappa}{4} \sum_{[i,j,k,l]}\sigma_{i} \sigma_{j} \sigma_{k} \sigma_{l}
\end{eqnarray}
and observing that if the lattice can be tiled by
a cube configuration minimizing the individual $h_c$
then the ground state energy density is
$\epsilon_0 = min\;  h_c$.

This approach reveals that a layered ground state
with parallel layers of flipped spins 
perpendicular to one of the lattice axes and
arbitrary interlayer spacing is degenerate with
the ferromagnetic ground state for all $\kappa$. This might
have been expected from the flip symmetry of the 
hamiltonian itself. In addition, at $\kappa=0$ an extra
ground state corresponding to diagonal flipped planes 
appears.
 
In the mean field approximation the spins
are in effect replaced by the average site magnetizations.
The calculation
of the mean field free energy is an elaboration of the
zero temperature approach
in which the energy is decomposed into a sum of individual cube terms.
The next to nearest neighbour and plaquette interactions
in the Gonihedric model give
the total mean field
free energy as the sum of elementary cube free energies $\phi(m_{c})$, given by
\begin{eqnarray}
\phi{(m_{C})} &=&- \frac{\kappa}{2}\sum_{<i,j>\subset C} m_{i} m_{j} 
\nonumber \\
&+& \frac{\kappa}{4}  \sum_{<<i,j>>\subset C }m_{i} m_{j}  \nonumber \\
&-& \frac{1-\kappa}{4} \sum_{[i,j,k,l] \subset C}m_{i} m_{j} m_{k} m_{l} 
\nonumber \\
&+&
\frac{1}{16}
\sum_{i \subset C}[(1+m_{i})ln(1+m_{i}) \nonumber \\
&+& (1-m_{i})ln(1-m_{i})]
\end{eqnarray}
where $m_{C}$ is the set of the eight magnetizations of the elementary cube.
This gives a set of eight mean-field equations
\begin{equation}
\frac{\partial\phi(m_{C})}{ \partial m_{i}}_{(i=1 {\ldots} 8)} =0
\end{equation}
one for each corner of the cube. Numerical iteration of these
equations shows a single transition from
a paramagnetic high temperature state to a layered, or 
the equivalent ferromagnetic, low temperature state.
The $\beta_c$ determined in this fashion decreases
quite sharply with $\kappa$.

\section{MONTE-CARLO SIMULATIONS}

In order to see how the full theory tallies
with the zero-temperature and mean-field results
we carried out Monte-carlo simulations \cite{11} for various
$\kappa$ values and lattices
of size $10^3,12^3,15^3,18^3,20^3$ and $25^3$.
Periodic boundary conditions were imposed
in the three directions and three internal perpendicular planes
of spins fixed to be $+ 1$.
A simple Metropolis update was used because of the difficulty
in concocting a cluster algorithm for a Hamiltonian with
such complicated interaction terms. The program was tested
on the standard nearest neighbour Ising model and
the some of the parameters used in
the generalized Ising models of \cite{9} to ensure it was working.
We measured the usual thermodynamic quantities for the model:
the energy $E$, specific heat $C$,
(standard) magnetization $M$, susceptibility $\chi$
and various cumulants.

We can clearly see the second order nature of the peak
in the specific heat in Fig.1

\bigskip

\vspace{1.2in}
\begin{figure}[h]
\vspace{1.2in}
\includegraphics{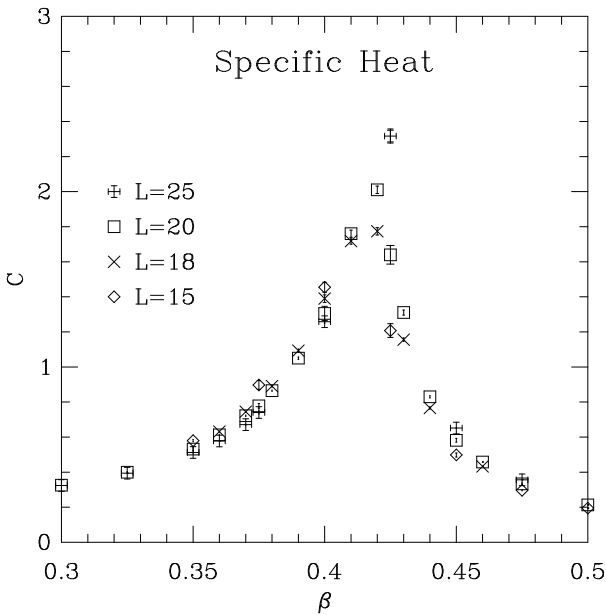}
\vspace{0.5in}
\caption{The specific Heat for $\kappa=1$ on
lattices of various size.}
\label{fig1:}
\end{figure}
Although extracting the exponent $\alpha$ from the scaling of the peak
is not particularly reliable because of the constant term that appears
in both the direct and finite-size scaling forms an approach
using the energy measurements gives $(\alpha -1 ) / \nu = -1.3(2)$.
Binder's magnetization cumulant and the scaling
of the susceptibility $\chi$ which is shown in Fig.2

\bigskip

\vspace{1.2in}
\begin{figure}[h]
\vspace{1.2in}
\includegraphics{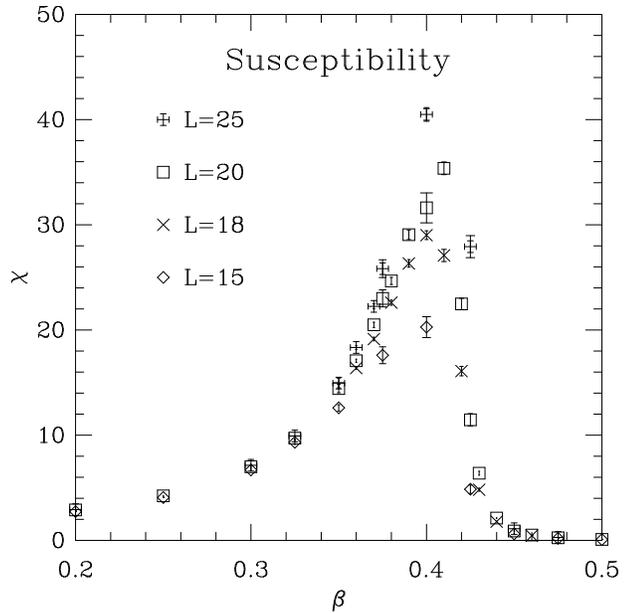}
\vspace{0.5in}
\caption{The susceptibility for $\kappa=1$ on
lattices of various size.}
\label{fig2:}
\end{figure}
\noindent
give an estimate of $\nu = 1.2(1)$. One also finds
$\gamma / \nu = 1.79(4)$ from the FSS of $\chi$.
All these exponents, rather remarkably given that the
model is defined in three dimensions, 
are close to the Onsager values of the two-dimensional
Ising model with nearest neighbour interactions.
The question of whether this is simply a numerical
coincidence or a true equivalence awaits 
an answer from further numerical work.
The exponents and the transition temperature
vary little, if at all, as $\kappa$ is increased.

Simulations of the $\kappa=0$ model \cite{12}
\begin{equation}
H= \frac{1}{2}\sum_{[i,j,k,l]}^{ }\sigma_{i} \sigma_{j}\sigma_{k} \sigma_{l}.
\label{e2}
\end{equation}
show that this
is a special case, displaying a first order
transition. The transition still appears
first order at $\kappa=0.1$ but softens rapidly
as $\kappa$ increases, so the crossover to the second
order behaviour seen at $\kappa=1$ is quite sharp.

\section{CONCLUSIONS}

We have investigated a class of Ising-like models
suggested by Savvidy and Wegner as a lattice discretization
of a particular random surface action that contains no bare surface
tension term. The models have a novel flip
symmetry that allows entire spin pnaes to be flipped at zero energy cost.
We have found
that the phase diagram displays only a single transition
to a low temperature layered state that, as a consequence
of the flip symmetry, is equivalent to a ferromagnetic
ground state.
For non-zero $\kappa$ the transition is second order,
whereas at, and probably close to, $\kappa=0$
a first order transition is manifest. 
The presence of some degree of self-avoidance, in the form
of a non-zero value for $\kappa$, would thus appear
to have an important influence on the universality
properties of the Savvidy-Wegner/gonihedric models.

It is also possible to define open surface
variants of the models described here, which
now incorporate link spins as well as vertex
spins, and to define models with
similar properties in higher dimensions.
All of these merit investigation both
from the string theory and statistical mechanical point of view.

\end{document}